\documentclass[cits]{PoS}
\input epsf 
\newcommand{\beq}{\begin{equation}}
\newcommand{\eeq}{\end{equation}}
\newcommand{\bea}{\begin{eqnarray}}
\newcommand{\eea}{\end{eqnarray}}

\newcommand{\ssc}{\scriptscriptstyle}

\newcommand{\hmu}{\hat{\mu}}
\newcommand{\tr}{{\rm tr}}

\newcommand{\vev}[1]{\Big\langle #1 \Big\rangle}
\newcommand{\bpsi}{\bar{\psi}}
\newcommand{\bK}{\bar{K}}

\title{Can Lorentz-breaking fermionic condensates form in large N strongly-coupled 
Lattice Gauge Theories?}

\ShortTitle{Can Lorentz-breaking fermionic condensates form in large N strongly-coupled 
LGT?}

\author{\speaker{E. T. Tomboulis}\\
        Univ of California, Los Angeles\\
        E-mail: \email{tombouli@physics.ucla.edu}}

\abstract{The possibility of Lorentz symmetry breaking (LSB) has attracted 
considerable attention in recent years for a variety of reasons, 
including the attractive prospect of the graviton as a Goldstone boson. 
Though a number of effective field theory analyses of such phenomena have 
recently been given it remains an open question whether they can take place 
in an underlying UV complete theory. Here we consider the question of 
LSB in large N lattice gauge theories in the strong coupling limit. 
We apply techniques that 
have previously been used to correctly predict the formation of chiral 
symmetry breaking condensates in this limit. 
Generalizing such methods to other composite operators we find that certain 
LSB condensates can indeed form. In addition, the interesting possibility 
arises of 
condensates that 'lock' internal with external symmetries. }

\FullConference{The XXVIII International Symposium on Lattice Field Theory, Lattice2010\\
		June 14-19, 2010\\
		Villasimius, Italy}

\begin{document}

The possibility of either explicit or spontaneous breaking of Lorentz symmetry has received considerable attention in recent years for a variety of phenomenological and theoretical reasons. 
The idea of spontaneous breaking goes back to Bjorken \cite{Bj} who proposed that the photon be interpreted as the Goldstone boson of such breaking; the same idea was naturally applied to the graviton \cite{PO} soon afterwards. In fact, a Goldstone graviton offers a rather attractive prospect for a quantum theory of gravity that evades the familiar difficulties of quantizing the metric field of General Relativity as an elementary field. This has been revived in recent years, and modern effective field 
theory treatments of the resulting Goldstone modes and their low energy interactions have been 
performed \cite{KT} - \cite{C}. Such analyses assume that Lorentz symmetry breaking occurs at some high (unification or Planck) scale and proceed to examine the low energy consequences. 
The central question then is whether such breaking can take place in an underlying 
theory which is UV complete. It indeed appears to be very difficult to come up with an UV healthy model where dynamical Lorentz breaking takes place at weak coupling. This may be just as well since 
this is naturally expected to be a strong-coupling dynamics phenomenon. Here we examine the question in $SU(N)$ or $U(N)$ lattice gauge theories in the strong coupling and large $N$ limits. 
This, as it is well known, is a model that gives a good qualitative depiction of all the 
basic non-perturbative features of QCD-like theories. We apply techniques that 
have previously been used to correctly predict the formation of chiral 
symmetry breaking condensates in this limit \cite{BBEG}, \cite{KS},  \cite{CJT}, 
generalizing such methods to other composite operators. We employ naive massless fermions, which 
automatically provide an anomaly-free, chirally invariant model,  and thus 
are well suited for our purposes since the doubling problem is irrelevant here - in fact, as it turns out,  the more degrees of freedom (color and flavor) the better.  We often write formulas for general dimension $d$ but are actually 
interested only in $d=4$.

The lattice action with naive massless fermions is given by
\beq
S= \sum_p \beta\,\tr U_p + \sum_{b=(x,\mu)}{1\over 2} \,\left[ \bar{\psi}(x) \gamma^\mu U_\mu(x) \psi(x+\hat{\mu}) - \bar{\psi}(x+\hat{\mu}) \gamma^\mu U^\dagger_\mu(x) \psi(x) \right]  \,.
\label{act1}
\eeq
We will be mainly concerned with expectations of the form $\bpsi(x)\Gamma^A\psi(x)$, where $\Gamma^A$ may stand for any of the Clifford algebra elements, such as
$\Gamma_S=1$, $\Gamma^\mu_V=\gamma^\mu$,  or $\Gamma^\mu_{A}=i\gamma^5\gamma^\mu $, or some other choice. Operators involving nearest neighbors (derivatives) will also be 
considered below. 
It is interesting to note that non-vanishing Lorentz-breaking condensates may also 
violate some discrete symmetries. Thus, for example, a non-vanishing vector condensate will also violate 
$C$, whereas an axial vector condensate will violate $P$, but a tensor condensate, as  in (\ref{Tfull2}) below, does not  violate either.


Since the operator $\bpsi(x)\Gamma^A\psi(x)$ is a fermion bilinear its expectation is related to the 
fermion 2-point function (full propagator) $G^{a,b}_{\alpha ,\beta }(x,y)=\vev{\psi^a_\alpha (x)
\bpsi^b_\beta(y)}$ in the limit $x=y$:
\beq
\vev{\bpsi(x)\Gamma^A \psi(x)}  =  -\tr \left[ G(x,x) \Gamma^A\right] 
 = -\tr_{\ssc D} \left[ \bar{G}(x,x) \Gamma^A\right] \label{exp1}
\eeq
with the second equality written explicitly in terms of the gauge invariant quantity 
$\bar{G} (x,x)\equiv \tr_{\ssc C} G(x,x)$. Here 
$\tr$ denotes trace over spinor and color (and flavor) indices, whereas $\tr_{\ssc C}$, $\tr_{\ssc D}$ 
denote traces over color and Dirac spinor indices, respectively.

To investigate such expectations we add to the action an external source $K^A$ which couples to $\bpsi(x)\Gamma^A\psi(x)$. One may more generally add a source for $G(x,x)$ of the form $K = \bar{K}{\bf 1_{\ssc C}}$, where ${\bf 1_{\ssc C}}$ 
denotes the unit matrix in color space and $\bK$ an arbitrary (invertible) matrix in spinor space. 
Coupling to a particular fermion bilinear then corresponds to a particular form of $\bK$;
e.g. $\bK=k n_\mu \gamma^\mu$, where $k$ an arbitrary number and $n_\mu$ an arbitrary unit vector, couples a source of magnitude $k$ and direction $n_\mu$ to $\bpsi\gamma^\mu \psi$. 

We write the action (\ref{act1}) in the presence of the external source more concisely in the form 
\beq 
S= \sum_p \beta\,\tr U_p + \sum_{x,y} \bpsi(x){\cal K}_{x,y}(U)\psi(y) \; ,\label{act2} 
\eeq
where 
\beq 
{\cal K}_{x,y}(U) = {\bf M}_{x,y}(U) + {\bf K}_{x,y}  \label{act3}
\eeq
with 
\beq
{\bf M}_{x,y}(U)  \equiv  {1\over 2}\,\left[ \gamma_\mu U_\mu(x)\, \delta_{y,x+\hmu} -
\gamma_\mu U^\dagger_\mu (x-\hmu)\,\delta_{y,x-\hmu} \right] \, , \qquad  
{\bf K}_{x,y}  \equiv  K\, \delta_{x,y} = \bK \,{\bf 1_{\ssc C}}\, \delta_{x,y} \;. \label{act3a}
\eeq
Note that ${\bf K}$ and ${\bf M}$ are matrices in spinor and color space as well as in lattice 
coordinate space. 

In the strong coupling limit $\beta\to 0$ the plaquette term in (\ref{act2}) is dropped. The 
corrections due to this term can be computed within the strong coupling cluster expansion, which, for sufficiently small $\beta$, converges. Hence they do not produce any qualitative change in the behavior obtained below at $\beta\to 0$. 
Setting $\beta=0$ in (\ref{act2}) then, $G(x,x)$  is given by 
\bea
G(x,x) & = & {1\over \int [DU] \, {\rm Det} {\cal K}(U)}
\int [DU] \,  {\rm Det}{\cal K}(U)\,
{\cal K}^{-1}_{x,x}(U)
\label{exp2a}\\
& = & {
\int [DU] \,  {\rm Det} [{\bf 1 + K^{-1} M}(U)]  \; 
   \left[[{\bf 1 + K^{-1} M}(U)]^{-1} {\bf K}^{-1}\right]_{x,x}
\over \int [DU] \, {\rm Det} [{\bf 1 + K^{-1} M}(U)] }  \,,    \label{exp2b}
\eea
from which the expectation of $\bpsi(x)\Gamma^A\psi(x)$ in the presence of the source is 
obtained from (\ref{exp1}). 

We evaluate (\ref{exp2b}) in the hopping expansion. This amounts to expanding (\ref{exp2b}) treating 
 ${\bf M}$ as the interaction and ${\bf K}$ as defining the inverse `bare propagator':
${\bf K}^{-1}_{x,y} = \bK^{-1} {\bf 1_{\ssc C}}\, \delta_{x,y}$.
The textbook version of the expansion is the case 
when the source is a mass term, i.e. $\bK=m{\bf 1_{\ssc D}}$.  
 Note that  ${\bf K}$ is purely local, whereas ${\bf M}$ has only 
nearest-neighbor non-vanishing elements ${\bf M}_{x, x+\hmu} =  {1\over 2}\gamma_\mu U_\mu(x) $ and 
${\bf M}_{x, x-\hmu} = {1\over 2}\gamma_\mu U^\dagger_\mu(x-\hmu)$.  
In the absence of the plaquette term integration over the gauge field results into non-vanishing contributions only if at least two $M$ factors with 
equal (mod $N$) number of $U$ and $U^\dagger$'s 
occur on each bond. 

The  expansion of the ${\cal K}^{-1}_{x,x}(U)$ is represented by all paths starting and ending at $x$, whereas that of 
the ${\rm Det} {\cal K}(U)$ by all closed paths \cite{R}.  Consistent with the above constraint on each bond resulting from the $U$-integrations the connected graphs giving the expectation 
(\ref{exp2b}) naturally fall into two classes: `tree graphs' and `loop graphs'. 
The tree graphs consist of paths starting and ending at $x$ and 
enclosing zero area (cf. graphs on the l.h.s. in Fig.1). 
We note in passing the well-known fact (see e.g. \cite{R}) concerning the hopping expansion that there are no restrictions on how many times a bond is revisited in drawing all such possible 
connected graphs.

Now, the set of tree graphs are the leading contribution in $N$. Loop graphs are down by powers 
of $1/N$ relative to tree graphs \cite{BBEG}. Thus, the set of tree graphs in the hopping  
expansion give the 
large $N$ limit of the theory. The sum of all tree graphs attached at site $x$ then 
constitute the full propagator $G(x,x)$ in this limit. 

The lowest order contribution is just the bare propagator ${\bf K}^{-1}_{x,x}$. 
Next consider trees extending only to the nearest neighbor (nn) sites. 
The simplest  such tree has only one `trunk' extending 
to any one of the nn-sites to $x$.  But one also has nearest-neighbor trees with $n$ trunks, where  each trunk extends only to any one of the $2d$ nn sites. Starting from these nearest neighbor trees, 
the full set of trees attached to the point $x$ can now be grouped as follows \cite{BBEG}.  
One simply observes that the full set of  `1-trunk' trees at $x$ is obtained by attaching to every 
1-trunk nn-tree at $x$ all possible trees at the site $x+\hmu$.  But the set of all trees attached at $x+\hmu$ comprise the full propagator 
$G(x+\hmu,x+\hmu)$. 
Similarly, the full set of `n-trunk' trees at $x$ is generated by attaching to every 
n-trunk nn-tree at $x$ all possible trees at each site 
$x+\hmu_j$, i.e. the full propagator $G(x+\hmu_j,x+\hmu_j)$, for $j=1,\ldots, n$.  Fig.1 
represents this  diagrammatically for $n=2$ and $n=3$. 
\begin{figure}
\includegraphics[width=\textwidth]{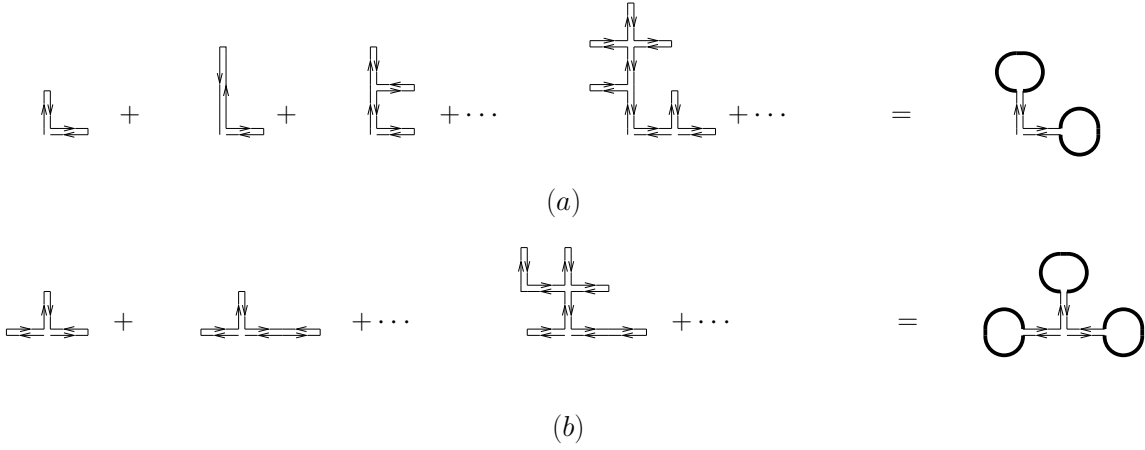}
\vspace{-15.5cm} \label{fig1}
\caption{Sum of: (a) 2-trunk; and (b) 3-trunk trees} \vspace{0.3cm}
\end{figure}
The set of all trees is now recovered by summing over all full n-trunk trees including the zeroth-order $n=0$ (no bottom-trunk), i.e. the bare propagator term. The resulting equation, depicted graphically in Fig.2, provides now a self-consistent equation for $G(x,x)$ in the large $N$ limit. 
\begin{figure}
\label{fig2}
\hspace{3cm}\includegraphics[width=0.6\textwidth]{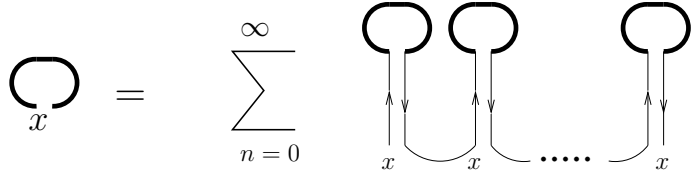}
\vspace{-10cm}  
\caption{The self-consistent equation for the sum of trees attached to site $x$, i.e. $G(x,x)$}
\end{figure}

With constant (position-independent) source $K$, translation invariance implies that $G(x,x)$ is in fact $x$-independent.  
Using the explicit expressions (\ref{act3a}), and bare propagator given by $\bK^{-1} {\bf 1_{\ssc C}}$, the r.h.s. in this equation may be easily evaluated. One finds 
\bea
G & = & \left[ {\bf 1}_{\ssc C} + \sum_{n=1}^{\infty} \left[ {(-1)\over 2}\, \bK^{-1} \,
\gamma^\mu G \gamma_\mu \right]^n \right] \bK^{-1} 
\label{Gfulla}\\
    & = & \left[ \bK {\bf 1}_{\ssc C} + {1\over 2}\, \gamma^\mu G \gamma_\mu \right]^{-1} \, . \label{Gfullb} 
\eea 
The hopping expansion which, in the large $N$ limit, gave (\ref{Gfulla}) converges for sufficiently large $||K||$. The resumed expression (\ref{Gfullb}), however, can be continued for all $K$. In particular, one is interested in possible solutions to (\ref{Gfullb}) for $K\to 0$. 

(\ref{Gfullb}) was obtained by resummation of the (infinite) set of leading graphs in the large $N$ limit. An alternative approach is the direct construction of the corresponding effective action defined 
as the Legendre transform $\hat{\Gamma}(G)$ of the free-energy w.r.t. the source $K$. Since here we deal with composite, viz. bilinear fermion operators, this is the effective action for composite operators \cite{CJT}. 
It is in fact quite straightforward to apply the general expression for the effective action given in 
\cite{CJT} to the theory (\ref{act2}) in the strong coupling limit. Variation of the resulting 
effective action $\hat{\Gamma}(G)$ again yields  (\ref{Gfullb}), as expected. This is in fact a rather more efficient and elegant way of arriving at the result: only {\it one} 2-PI graph need be evaluated in the large $N$ limit. 

We also note in passing that still another approach is that proposed in \cite{KS}. This approach, however, is mathematically inherently ambiguous and care must be exercised in applying it. 
If this is done it gives the same qualitative results but in a much lengthier and inefficient manner. 

It is now easy to examine particular solutions of (\ref{Gfullb}) that are picked out by 
appropriate choice of the source $K$. (We will not examine here the most general solution.) 
In all cases at large $||K||$ the solution reproduces, of course, the perturbative hopping expansion 
solution. We are, however, interested in the vanishing-source limit.  
In the case of scalar source, the solution is $G=g_S(K) {\bf 1_{\ssc D}} 
{\bf 1_{\ssc C}}$, and one finds $g_S(0)= \sqrt{{2\over d}}$. For the scalar condensate one thus gets 
\beq
\vev{\bpsi(x)\psi(x)} = -NS \sqrt{{2\over d}} \, ,\label{Scond}
\eeq
where $S=\tr {\bf 1_{\ssc D}}$ is the number of spinor components. This  reproduces the result in 
\cite{BBEG}. For vector or axial vector source $K=k (\Gamma \cdot n){\bf 1_{\ssc C}}$, 
the solution is of the form $G= g(k)  (\Gamma \cdot n)^{-1}{\bf 1_{\ssc C}}$, where 
$\Gamma^\mu$ stands for either  $ \Gamma_V^\mu$ or $\Gamma_A^\mu$.   
(Note that in either case one has $(\Gamma \cdot n)^{-1} =(\Gamma\cdot n)$.) One now finds 
$g_V(0)= i \sqrt{ 2/ (d-2)}$ whereas $g_A(0)= \sqrt{ 2/(d-2)}$. For the axial vector 
condensate one thus gets 
\beq
\vev{\bpsi(x)\,i\gamma^5\gamma^\mu \,\psi(x)} = -NS \sqrt{{2\over (d-2)}}\, n^\mu \,.
\label{AVcond}
\eeq
In the vector case, however, the resulting expectation is imaginary. Indeed, the solution turns complex for small source $K$. This would seem to indicate that no vector condensate actually forms.
But this does not mean that other condensates induced in the presence of a vector source do not survive as the source is turned of. Consider the operator $\bpsi(x)\gamma^\mu \sigma_{\kappa\lambda}\psi(x)$, where 
$\sigma_{\kappa\lambda}={ i\over 2} [\gamma_\kappa, \gamma_\lambda]$. This condensate, which is of interest for LSB-induced gravity theories, is also induced in the presence of a 
vector source. In this case in the vanishing-source limit one gets 
\beq
\vev{\bpsi(x)\gamma^\mu \sigma_{\kappa\lambda}\psi(x)}=  NS \sqrt{{2\over (d-2)}}\;
[\,g^\mu_\kappa n_\lambda - g^\mu_\lambda n_\kappa\,] \,. \label{Ccond}
\eeq

Other chiral or Lorentz symmetry-breaking condensates involving more complicated operators such as 
lattice nearest-neighbor (continuum derivative) couplings may also be induced. 
The (gauge invariant) operator 
 \beq
  O_{\nu\mu}(x) \equiv  \bpsi(x)\gamma_\nu U_\mu(x) \psi(x+\hmu) - 
\bpsi(x)\gamma_\nu U^\dagger_\mu(x-\hmu) \psi(x-\hmu) \, , \label{Tcond}
\eeq
in particular, is of special interest. In its continuum limit it corresponds to 
the tensor operator $\bpsi(x)\gamma^\nu\partial_\mu\psi(x)$, for which a non-vanishing 
condensate is a natural starting point for a theory of the graviton as a Goldstone boson \cite{KT}. 
In the presence of a vector (or axial vector) source (\ref{Tcond}) acquires a non-zero expectation. 
The full expectation is obtained by attaching the full set of trees, i.e. full propagators $G$ at 
the sites $x$ and $x+\hmu$ giving the graph shown in Fig.3. This is easily evaluated and in the limit of vanishing source yields 
\beq
\vev{O_{\nu\mu}}=   -{1\over (d-2)} NS \left[ 2n_\nu n_\mu - g_{\nu\mu}\right] \, . \label{Tfull2}
\eeq
(\ref{Tfull2}) is a non-vanishing tensorial condensate {\it not} proportional to $g_{\nu\mu}$, i.e. 
an $SO(4)$-breaking (Lorentz-breaking) condensate. (A tensorial condensate proportional to the 
metric tensor is not Lorentz-breaking.) 
Different patterns of breaking, partial or complete, can be obtained by including fermions 
$\psi^i(x)$ of different flavor $i$ coupled to vector sources 
of different orientation $n_i^\mu$. If $N_F$ flavors are present (\ref{Tfull2}) becomes 
\beq
\vev{O_{\nu\mu}}=   {1\over 2} NN_FS \left[ \,g_{\nu\mu} - 
{2 \over N_F} \sum_i   n^i_\nu n^i_\mu \right] \, . \label{Tfull3}
\eeq 
The strongly coupled 
lattice model considered here provides in fact an explicit realization of the scenario envisioned 
in \cite{KT}. One may, for contrast, also consider the effect of the scalar condensate on (\ref{Tcond}). Repeating the calculation with a scalar source replacing the vector source  one now 
gets that $\vev{O_{\nu\mu}}$ is proportional to $g_{\nu\mu}$. Thus, as expected, no Lorentz symmetry breaking is induced in this case. 
\begin{figure}
\hspace{3cm}\includegraphics[width=0.7\textwidth]{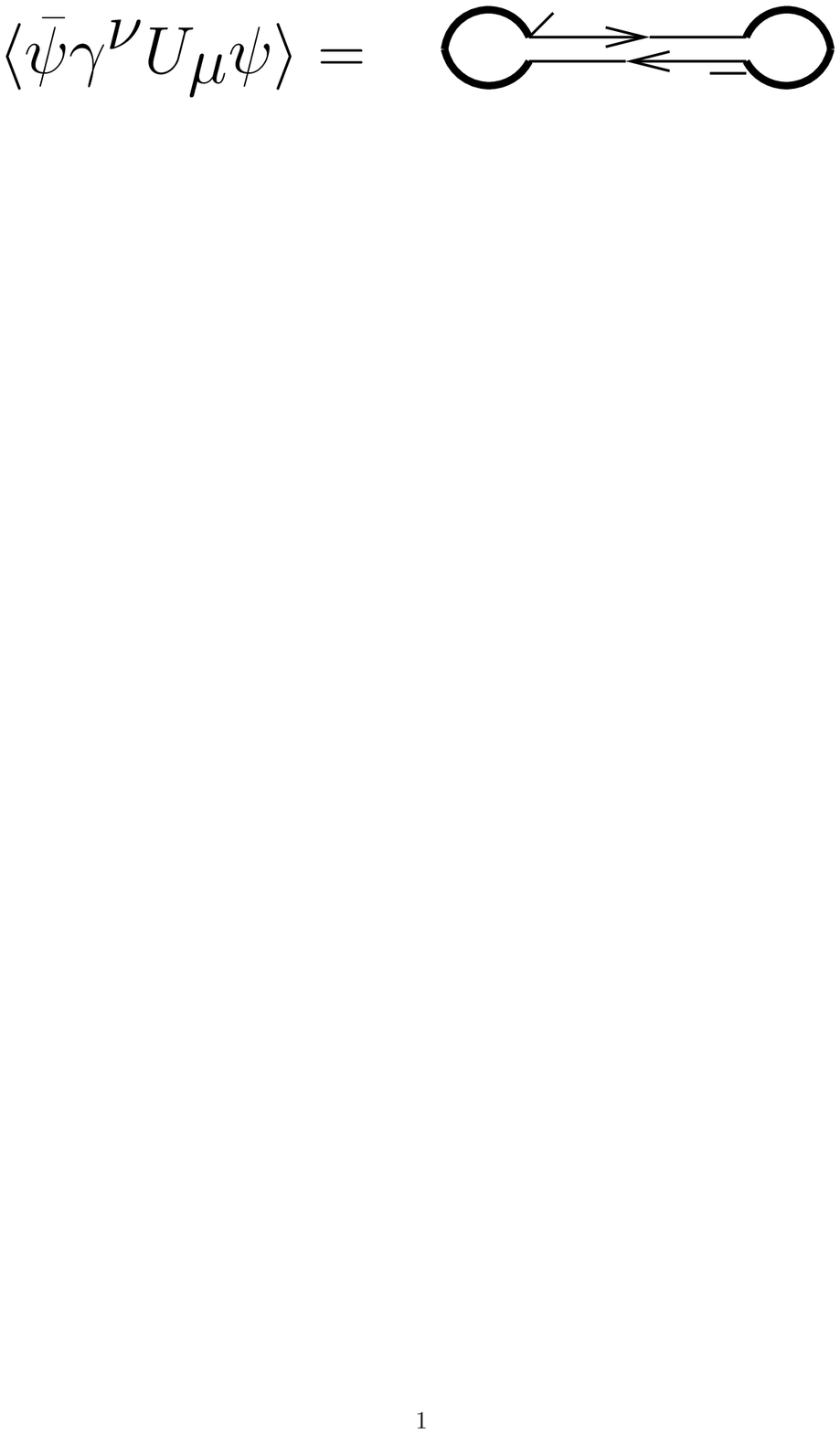}
\vspace{-17cm} \label{fig3}
\caption{Graph for the expectation after attaching full trees; the short lines denote the different 
directions of the $\gamma^\nu$, $\gamma^\mu$ factors in it. }
\end{figure}

When  internal (global) symmetry groups are present, 
a further possibility arises, i.e. condensate formation that 'locks'  space-time and internal symmetries. This possibility can be 
equally well explored within our strong coupling lattice gauge models. 

The most straightforward example is provided by taking the internal symmetry to be a copy 
of the (Euclidean) space-time symmetry, i.e. an internal $SO(d)$ group with the fermions transforming as Dirac spinors under it.  Denoting the gamma matrices acting on the internal space by 
$\gamma^m$, consider the  operator
$\bpsi(x) \gamma^n\,(i\gamma_5\gamma_\nu)\psi(x)$ involving an internal vector and an external axial vector. Non-vanishing vev's of such fermion bilinears can lead to locking between the corresponding groups. To compute such a vev we again first 
introduce appropriate sources.  
Different fermion flavors can be coupled to different sources. In $d=4$ take the number of 
flavors to be (a multiple of) four coupled to corresponding sources along the elements $n_{(i)}^\mu$ and $n_{(i)}^m$, $i=1, \ldots, 4$, of an orthonormal tetrad set in external and internal space, respectively. Proceeding as before, either by tree resummation, or, more efficiently, by direct  construction of the effective action, one now, in the limit of vanishing sources, obtains the result
\beq
\vev{\bpsi(x) \gamma^n\,(i\gamma_5\gamma_\nu)\psi(x)} = -  
NS^2 \sum_i  n_{(i)}^n n_{(i)\,\nu} = - NS^2 \delta^n_\nu  \,. \label{CLva}
\eeq
(\ref{CLva}) represents complete locking of the internal and external symmetry, i.e. breaking to the diagonal $SO(4)$ subgroup of the original symmetry: $SO(4) \times 
SO(4)_I \to SO(4)_D$. The condensate remains invariant only under simultaneous equal internal and external rotations. 

The obvious question arises: how can such locking work in Minkowski space? 
There appear to be two possible choices. 
One choice is the standard Wick rotation where the external group gets decompactified to $SO(3,1)$ whereas the internal group remains compact. 
The condensate (\ref{CLva}) is now invariant only under simultaneous $SO(3)$ (spatial) rotations, i.e. 
$SO(3,1)\times SO(4)_I \to SO(3)_D$. 
The second possibility is to define the passage to Minkowski space to also involve a `Wick rotation' 
of the internal group decompactifying it. Full locking then is preserved, i..e (\ref{CLva}) remains invariant under $SO(3,1)_D$. The obvious difficulty now is that an internal non-compact group, such as $SO(3,1)$, possesses only non-unitary finite-dimensional representations. This, of course, leads in general to unitarity violation. 
The only way out is to take the fermions to transform under a unitary, i.e. an infinite dimensional representation of the internal non-compact group. For an internal group the usual formalism applies whether one uses finite or infinite dimensional unitary representations.\footnote{Thus the 
problems of physical interpretation with respect to particle spectrum and spin-statistics that plague 
the use of infinite dimensional representations for external (Lorentz) groups are not relevant in this 
context.} The new feature implied by the use of an infinite dimensional representation is the infinite number of components associated with the internal group index. This locking mechanism may offer a novel approach to a quantum gravity theory. At any rate, it would be interesting to work out the effective field theory for it at low energies.

\vspace{0.5cm}
This research was partially supported by NSF-PHY-0852438.


\end{document}